\documentclass[letterpaper, 10pt, conference,pdftex]{ieeeconf}  %

\IEEEoverridecommandlockouts                              %

\usepackage{amsmath} %
\usepackage{amssymb}  %
\usepackage{newtxtext,newtxmath}
\usepackage{graphicx}
\graphicspath{{figures/}}
\usepackage{algorithm}
\usepackage{algorithmic}
\usepackage{url}
\usepackage{siunitx}
\usepackage{etoolbox}%
\newcommand{\ubold}{\fontseries{b}\selectfont}%
\robustify\ubold%
\usepackage{optidef}
\usepackage{color}
\usepackage[labelformat=simple]{subcaption}
\usepackage{soul}

\usepackage{bm}

\newcommand{\sR}{\mathbb{R}}

\newcommand{\ths}{\theta^{\star}}
\newcommand{\Ps}{P^{\star}}
\newcommand{\Kxe}{\hat{K}_{\hat{x}}}
\newcommand{\Kde}{\hat{K}_{d}}
\newcommand{\thhat}{\hat{\theta}}
\newcommand{\pob}{\bm{p}_{\mathrm{ob}}}

\newcommand{\rv}[1]{{#1}}

\newtheorem{remark}{Remark}

\DeclareMathOperator*{\argmin}{arg\,min}

\title{\LARGE \bf
Direct closed-loop identification of continuous-time systems\\
 using fixed-pole observer model*
}

\author{Ichiro Maruta$^{1}$ and Toshiharu Sugie$^{2}$%
\thanks{*This work is supported by JSPS KAKENHI Grant numbers  JP20H02170 and JP17H03281.}
\thanks{$^{1}$ Department of Aeronautics and Astronautics, Engineering,
        Kyoto University, Kyoto, 615-8540, Japan.
        {\tt\small maruta@kuaero.kyoto-u.ac.jp}}
\thanks{$^{2}$ Komatsu CRC, Graduate School of Engineering, Osaka University, %
Suita, Osaka 565-0871, Japan.
        {\tt\small  sugie@jrl.eng.osaka-u.ac.jp}. }
\thanks{© 2022 IEEE.  Personal use of this material is permitted.  Permission from IEEE must be obtained for all other uses, in any current or future media, including reprinting/republishing this material for advertising or promotional purposes, creating new collective works, for resale or redistribution to servers or lists, or reuse of any copyrighted component of this work in other works.}}
 
\begin{document}

\maketitle
\thispagestyle{empty}
\pagestyle{empty}

\begin{abstract}
This paper provides a method for obtaining a continuous-time model of a target system in closed-loop from input-output data alone, \rv{in the case where no knowledge of the controllers nor excitation signals is available and I/O data may suffer from unknown offsets}. The proposed method is based on a fixed-pole observer model, which 
is a reasonable continuous-time version corresponding to the innovation model in discrete-time and 
allows the identification of unstable target systems.
 \rv{Furthermore, it is shown that the proposed method can be attributed to a convex optimization problem by fixing the observer poles.}
The method is within the framework of the stabilized output error method and shares usability advantages such as robustness to noise with complex dynamics and applicability to a wide class of models.
The effectiveness of the method is illustrated through numerical examples.
\end{abstract}

\section{INTRODUCTION}

System modeling is the foundation of control system design, and among them, closed-loop identification is often necessary for practical use. 
When the target system is open-loop unstable, closed-loop identification must be performed after stabilizing the system with some feedback. 
Furthermore, in large-scale networked systems, which have been attracting much attention in recent years, each subsystem must be identified within the network, and closed-loop identification is inevitable because of the presence of feedback from other subsystems. 
In this case, it is not possible to obtain accurate information on the feedback part. 
An identification method that does not require any controller information is desirable to deal with such a case. 
In addition, though most identification methods handle discrete-time models, it is often desired to obtain continuous-time models because they are intuitive to most control engineers and the system parameters are consistent with physical properties.  
 
In the conventional closed-loop identification method within the framework of the Prediction Error Method, various “direct methods” have been developed that do not require any information on the controller
(see surveys \cite{FL99,VS95}).
In these methods, accurate modeling of the observed noise characteristics is important in general, but it is difficult to accurately model the noise characteristics in the real world due to its complexity. 
In addition, although it may not be widely recognized, it is not easy to identify open-loop unstable systems \cite{FL00}. 
On the other hand, in the framework of subspace identification methods, SSARX \cite{Jan03} and PBSID \cite{CP05} are well known that do not require controller information and can handle unstable systems. 
The effectiveness of these methods has also been pointed out in a survey paper \cite{VdV13}. 
Though they may not work well in the presence of complex colored noise, a method which can overcome such shortcomings has been recently proposed \cite{MS22}. 
What these methods have in common is that they employ an observer (or Kalman filter) form called the innovation form, which makes it easy to deal with unstable systems. However, all of these papers discuss discrete-time systems only. 
Hence, the data sampling time must be chosen very carefully, and we have to convert the obtained discrete-time model into a continuous-time one. 
Consequently, it is often not easy to obtain an accurate continuous-time model by these methods.

On the other hand, various methods to obtain a continuous-time model directly from input-output data have been actively developed (see  \cite{GW08,Gan15}). 
A few papers (e.g., \cite{GGYH08, MS13}) discuss closed-loop identification, but they cannot handle unstable systems without controller knowledge. One exception is the stabilized output error (OE) method proposed by the authors \cite{MS21}. 
The method of \cite{MS21} is applicable if one can find a controller that stabilizes the target system, where 
the controller can be different from the original one in the closed-loop. 
Also, its effectiveness has been demonstrated through various numerical examples. 
However, it is not clear how to find an appropriate stabilizing controller for the system to be identified. 

This paper aims to give a method to directly identify a continuous-time model using only input/output data of a system in an uncertain closed-loop environment. Specifically, we give a method for handling the case where the controller is completely unknown in the stabilized OE method framework and propose a fixed-pole observer model,
which is a reasonable continuous-time version corresponding to the innovation model in discrete-time.
Furthermore, many variants can be created based on the idea of the proposed method, and here we propose a fixed-pole extended observer model. It can handle time-series data with non-stationary trends, similar to the ARIX and ARIMAX models for discrete-time systems. 
\rv{Finally, the effectiveness of the proposed method will be evaluated through detailed simulation results.}

In this paper, for a vector $x$, $x_i$ represents its $i$-th element, and $\mathcal{N}\left(\mu,\sigma^2\right)$ denotes the normal distribution with mean $\mu$ and variance $\sigma^2$.
\section{PROBLEM SETTING}\label{sec:prob}

\begin{figure}[ht]
 \centering
 \includegraphics[scale=1]{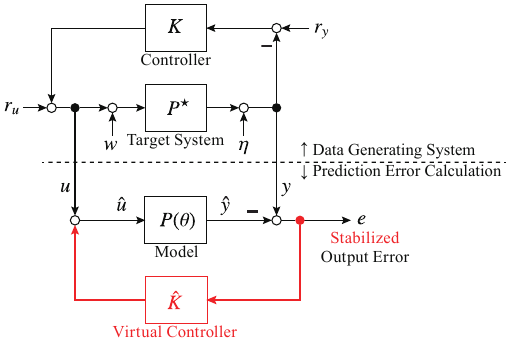}
 \caption{Data generating system and stabilized output error method}
 \label{fig:datagen}
\end{figure}

We consider a closed-loop system described by
the differential equation
\begin{align}
y(t)  &= \Ps(p) \left(u(t) +w(t)\right) + \eta(t), \\
u(t)  &= r_u(t) + K(p) \left(r_y(t)-y(t)\right),  
\end{align}
where $p$ denotes the time-domain differential operator and $\Ps(p)$ is the rational function of $p$ which represents the plant to be identified, and  $K(p)$ represents an unknown controller which stabilizes $\Ps(p)$ (see the upper half of Fig.~\ref{fig:datagen}). 
The signals $u(t)$, $y(t)$, $r_u(t)$ and $r_y(t)$ are the plant input, the plant output measurement, the excitation signal and the reference signal, respectively.
$r_u(t)$ and $r_y(t)$ may be unknown but are supposed to be rich enough for identification purpose.  
And $w(t)$ and $\eta(t)$ denote the input disturbance and the measurement noise, respectively,
which may be colored.

The plant is assumed to belong to a known parametrized set
\[
{\cal P} := \{ P(p, \theta) ~|~ \theta \in \sR^{n_{\theta}} \},
\]
so it is described  by $\Ps(p)=P(p, \ths)$ for some parameter $\ths\in \sR^{n_{\theta}}$. 
The I/O data are collected with the sampling time $h$ and given by
\begin{align}
{\cal  Z} =\left\{ u(kh), y(kh)  ~\middle|~ k=0,1,\cdots, N-1 \right\}
\end{align}
where $N$ is the data length. 

\rv{The sampling time is supposed to be very short (e.g., $\SI{1}{ms}$ $\sim$ $\SI{0.01}{ms}$ for a system with dynamics on a time scale of $\SI{1}{s}$) so that we can capture the continuous-time dynamics precisely in digital implementation.}

The problem here is to estimate $\ths$ from the I/O data alone.

\section{STABILIZED OUTPUT ERROR  METHODS}\label{sec:soe}

As a preliminary, we briefly outline the stabilized OE method framework.

\subsection{Output Error (OE) Method}

Most natural way to identify $P(p, \ths)$ based on I/O data would be to employ the plant model described by 
\begin{align}
\hat{y} (t) & = P(p, \theta) u(t)
\end{align} 
and obtain the parameter estimate $\thhat$ via 
\begin{align}
\thhat & = \argmin_{\theta} \sum_{k=0}^{N-1} e^2(hk)\label{eq:opt} \\
e(t)   & := y(t) - \hat{y} (t) \label{eq:e}
\end{align}

The lower half of Fig.~\ref{fig:datagen}, excluding the virtual controller (in red), corresponds to the calculation of this output error.
This is nothing but the output error method. 
In applications to closed-loop system identification, the estimate is biased by the correlation between $u$ and noise, but is often acceptable if the signal-to-noise ratio is high and the target plant $\Ps$ is stable.
However, when $\Ps$ is unstable,
this method does not work well. Because when $\theta$ takes the desired value $\ths$, the prediction system becomes unstable and the output error $e(t)$ diverges.

\subsection{Stabilized OE Method}

Hence,  \cite{MS21} has proposed the stabilized OE method, which is 
to employ the following form;
\begin{align}
\hat{y} (t) & = P(p, \theta) \hat{u}(t),   &
\hat{u}(t)  & = u(t) + \hat{K}(p) e(t), 
\end{align}
and (\ref{eq:e}), where $\hat{K}(p)$ is called a virtual controller which stabilizes $P(p, \ths)$. 
Introduction of the virtual controller with error feedback 
plays an essential role to handle unstable plants \cite{MS21}. 
When  $K(p)$ is known,  it has been shown that $\hat{K}(p) = K(p)$ produces an unbiased  
plant model. More importantly, this method turns out to be quite insensitive to the discrepancy 
between $\hat{K}(p)$ and $K(p)$ as demonstrated in \cite{MS21}. 
One open problem here is how to find an appropriate virtual
controller when we have no clue about $K(p)$ or $P(p, \ths)$.

\section{\rv{PROPOSED METHOD}}

Now, we give a new scheme to solve the above issue. As a variant of the stabilized OE method, we will adopt an observer for $P(p, \theta)$ instead of the output feedback virtual controller 
$\hat{K}(p)$.

\subsection{\rv{Fixed-pole observer model}}

Let $A(\theta) \in \sR^{n \times n}$, 
$B(\theta) \in \sR^{n \times 1}$ and 
$C(\theta) \in \sR^{1 \times n}$
be matrices of a minimal state space representation, 
and $a(p,\theta)$ and $b(p,\theta)$ be the denominator 
and the numerator of $P(p, \theta)$, namely,
\begin{align}
P(p,\theta)=C(\theta) \left(p I - A(\theta)\right)^{-1}B(\theta) %
= \frac{b(p,\theta)}{a(p,\theta)}.%
\end{align}
Without loss of generality, the denominator is assumed to be monic, so
\begin{align}
a(p, \theta) = \left| pI - A(\theta) \right|
\end{align}
holds, where $\left| ~ \cdot ~ \right|$ denotes the determinant of the matrix.
In order to stabilize $P(p,\theta)$, we will employ a virtual controller, which feedback $e(t)$ to the model state $\hat{x}$ and forms the observer 
\begin{subequations}\label{eq:ob}
\begin{align}
\dot{\hat{x}}(t) &=A(\theta) \hat{x}(t) +B(\theta) u(t) + \Kxe (\theta)e(t) \\
\hat{y}(t) &= C(\theta) \hat{x}(t) \label{eq:ob:y}\\ %
e(t) &= y(t) -\hat{y}(t). \label{eq:ob:e}
\end{align}
\end{subequations}
First, we will derive the I/O relation between $e(t)$ and  $\{u(t)$,~$y(t) \}$. From the above equations, we have
\begin{align}
e(t) & = G_{ey} (p,\theta) y(t) - G_{eu} (p,\theta) u(t) \label{eq:e:gey:geu}\\
G_{ey} (p,\theta) & : = (1 -C (\theta)(pI-A (\theta)+\Kxe(\theta) C (\theta))^{-1} \Kxe(\theta)  \\
G_{eu} (p,\theta) & := C (\theta) (pI-A (\theta)+\Kxe (\theta)C (\theta))^{-1}B (\theta).
\end{align}
The denominator of $G_{ey}(p, \theta)$ is given by
\begin{align}
\phi(s, \theta) & = \left| sI - A(\theta) +\Kxe(\theta)C(\theta) \right|
\end{align}
which is also the denominator $G_{eu}(p, \theta)$. On the other hand, 
 the numerator of $G_{ey} (p,\theta)$ turns out to be  $a(p,\theta)$. 
Because its inverse is given by
\begin{align}
G_{ey}^{-1}(s,\theta) =1  + C (\theta)(sI-A (\theta))^{-1} \Kxe (\theta)
\end{align}
and the denominator of the above is $a(p,\theta)$.
While, it is known  that  $G_{ey} (p,\theta)$ and $P(p,\theta)$ share the same numerator 
$b(p, \theta)$, because the poles of $P(p,\theta)$ can be changed  but the zeros 
are not affected by the observer gain. Hence, we obtain
\begin{align}
e(t)  &= \frac{a(s,\theta)}{\phi(s, \theta)} y(t) - \frac{b(s,\theta)}{\phi(s,\theta)} u(t).
\label{eq:error}
\end{align}

Based on the above observations,  we propose to  use the observer (\ref{eq:ob}) with the
gain $\Kxe (\theta)$ satisfying  
\begin{align}
\phi(p, \theta) = \phi^{*}(p) := \prod_{i=1}^n (p- \lambda_i),    ~~ \lambda_i \in \pob
\label{eq:phistar}
\end{align}
where $\pob$ denotes the set of desired poles which are chosen by the user.
Then, calculate $\hat{\theta}$ from (\ref{eq:opt}).

There are two important points. One is that we can always stabilize $P(p, \theta)$ regardless of
the value of $\theta$, because the observer gain is not fixed but can vary according to $\theta$. This is a big leap from using fixed virtual controllers in \cite{MS21}. 
The other is that the output error $e(t)$ is affine in $\theta$ from (\ref{eq:error}) and (\ref{eq:phistar}) as long as $a(p,\theta)$ and $b(p, \theta)$ are affine in $\theta$. In this case,  
 the optimization problem (\ref{eq:opt}) is convex.  This is a big advantage.

\begin{remark}
A straightforward approach is to search for $\Kxe$ simultaneously with $\theta$ by minimizing the output error (\ref{eq:opt}). In fact, the above observer model (with parameters $\Kxe$ and $\theta$) in discrete-time is called the innovation model, and the optimization of (\ref{eq:opt}) over both $\Kxe$ and $\theta$ yields the Kalman filter.
However, for the continuous-time model, there is a trivial solution that takes $\Kxe$ extremely large and makes $e(t) \to 0$ regardless of $\theta$, and this approach does not work.
\end{remark}

\subsection{\rv{Fixed-pole extended observer model}}\label{sec:proposed:exo}
While there are many possible variations on how to provide stability,
we introduce one noteworthy and useful variant.

Here we configure the virtual controller as follows so that the prediction system is an extended observer
\begin{subequations}
\begin{align}
\dot{\hat{x}}(t) &=  A(\theta)\hat{x}(t) + B(\theta)\left(u(t)+d(t)\right) + \Kxe(\theta)e(t)\\
\dot{d}(t) &= \Kde(\theta) e(t)  %
\end{align}
\end{subequations}
along with (\ref{eq:ob:y}) and (\ref{eq:ob:e}).  Here, $d(t)$ is introduced as an input correction term and $\Kde\in\sR$ is the corresponding observer gain.
From the above, 
\begin{align}
e(t) &= G_{ey}(p,\theta) y(t) - G_{eu}(p,\theta)(u(t) +d(t))  \\
d(t) & = \frac{\Kde(\theta)}{p} e(t) 
\end{align}
hold. With aid of (\ref{eq:e:gey:geu}) and (\ref{eq:error}), we have
\begin{align}
e(t) %
   & =  \frac{a(p,\theta) p}{\phi_{ext}(p,\theta)} y(t) - 
\frac{b(p,\theta) p}{\phi_{ext}(p,\theta)} u(t)  \label{eq:obext} \\
  \phi_{ext}(p,\theta)& := \phi(p,\theta) p + \Kde(\theta) b(s,\theta)
\end{align}
By choosing  the observer gain 
$\left[\Kxe(\theta)^\top, \Kde(\theta)^\top\right]^\top$
so that
\begin{align}
\phi_{ext}(p, \theta) = \phi_{ext}^{*}(p) = \prod_{i=1}^{n+1} (p- \lambda_i),    ~~ \lambda_i \in \pob
\label{eq:phiextstar}
\end{align}
holds, the optimization problem (\ref{eq:opt}) becomes convex again.
Since the extended observer corresponds to the observer for an augmented system 
$\frac{1}{p} P(p,\theta)$,
we can always find such an observer gain satisfying (\ref{eq:phiextstar}) 
provided %
$P(p,\theta)$ does not have a zero at the origin. 
Note that from (\ref{eq:obext}), its is easy to see that the effect of unknown offset 
in both $y(t)$ and $u(t)$ vanishes asymptotically because of the differential operator $p$ in both numerators. 
The use of this model is illustrated through examples in the next section.

\section{SIMULATION}

In this section, two numerical examples are shown to demonstrate the effectiveness of the proposed method. 
In the examples, 100 trials based on different random noise realizations are performed to see the statistics of the results.
The Levenberg-Marquardt method, implemented in MATLAB as \texttt{lsqnonlin} function, was used in the optimization required by the proposed method.

\subsection{Identification in closed-loop systems with unknown controller}\label{sec:ex:maglev}

The first example is the closed-loop identification of an unstable plant and is based on the model of the actual magnetic levitation system in \cite{SSI1993}.
The data generating system is as shown in Fig.~\ref{fig:datagen}, and the systems are as follows:
\begin{align}
    P(p,\theta) &= \frac{\theta_4}{p^3 + \theta_1 p^2 + \theta_2 p + \theta_3},\\
    \ths &= [13.33, -494.4, -6593, 7148]^\top,\\
    K(p) &= \frac{\SI{1.197e5}{}(p+9.294)(p+13.99)(p+20.9)}{(p+399.9)(p+0.1)\left((p+121.5)^2+141.1^2\right)}.
\end{align}

\begin{figure}[htbp]
  \centering
  \includegraphics[scale=1.0]{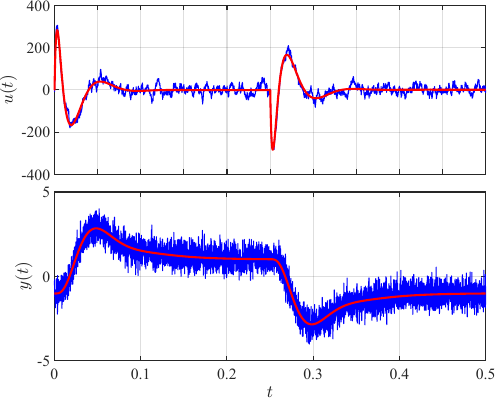}
  \caption{Example of input/output data for identification (Magnetic levitation system)}
  \label{fig:ex:maglev:io}
\end{figure}

An example of input/output data is shown in Fig.~\ref{fig:ex:maglev:io} as blue lines.
Here, the excitation signal is a square wave with a $\SI{0.5}{s}$ period applied to $r_y$.
The input/output data in the third period, when the effect of the initial state disappears, is used for system identification, and $5000 (=N)$ data are sampled at intervals of $\SI{0.1}{ms} (=h)$.

The disturbance $w(t)$ and the measurement noise $\eta(t)$ are zero-order hold signals of random numbers sampled every $\SI{0.1}{ms}$ from $\mathcal{N}(0,0.5^2)$, respectively.
To visualize the signal-to-noise ratio, the noise-free signals are shown as red lines in Fig.~\ref{fig:ex:maglev:io}.

\begin{figure}[htbp]
  \centering
  \begin{minipage}[b]{0.99\linewidth}
  \centering  
  \includegraphics[scale=1]{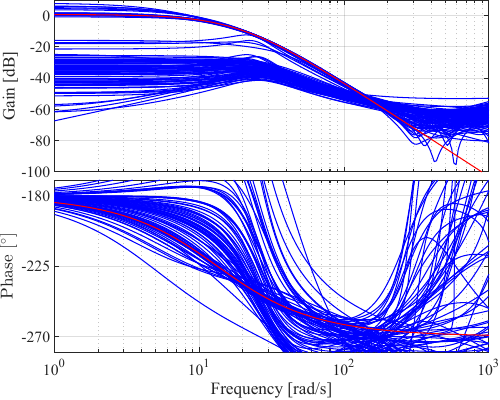}
  \subcaption{ARMAX}
  \label{fig:ex:maglev:bode:armax}
  \end{minipage}
  \\[2mm]
  \begin{minipage}[b]{0.99\linewidth}
  \centering  
  \includegraphics[scale=1]{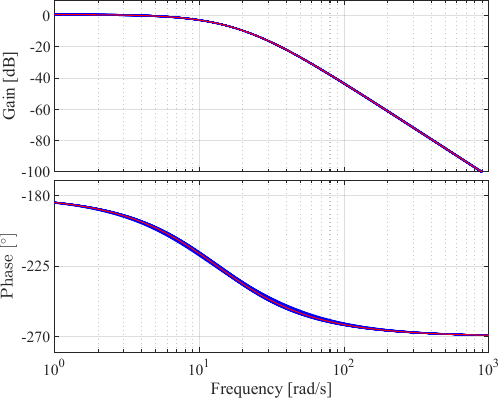}
  \subcaption{Stabilized OE (fixed-pole observer model)}
   \label{fig:ex:maglev:soem_ob}
  \end{minipage}
  \caption{Frequency response of model}
  \label{fig:ex:maglev:bode}
\end{figure}

Well-established methods for the closed-loop identification problem of such unstable systems include prediction error methods based on discrete-time prediction models such as ARX, ARMAX, and Box-Jenkins models\footnote{\rv{Another possible model will be error-in-variables model, but it has not been tried here.}}.
However, the system identification based on discrete-time models is prone to failure if the sampling interval and the dynamics of the target system do not match \cite{RG2002}.
This problem is especially likely to occur in closed-loop system identification, where the sampling interval is often selected based on the requirements of the control system design.
For reference, the results based on the ARMAX model, which worked best among ARX, ARMAX, and Box-Jenkins models, are shown in Fig.~\ref{fig:ex:maglev:bode:armax}.
In the figure, the frequency response of the target system is shown by the red line, and the blue lines show the characteristics of the model obtained over 100 trials based on different realizations of noise, respectively.
This result was obtained with the \texttt{armax} function of the MATLAB System Identification Toolbox, and although the choice of model order and other settings were adjusted to obtain the best results, the results are not good.
As seen in this example, there are a wide range of situations where system identification based on discrete-time predictive models is difficult.

In the case where the controller is not known at all, the method proposed in this paper can be applied. Fig.~\ref{fig:ex:maglev:soem_ob} shows the results for a  fixed-pole observer model with $\pob = \left\{-3, -3\pm i\right\}$, they are quite good. \rv{For comparison, we performed the identification by using the stabilized OE with ($\hat{K}(p) = K(p)$), where $K(p)$ is assumed to be perfectly known. The results are almost the same as Fig.}~\ref{fig:ex:maglev:soem_ob}\rv{, though the estimated bode plots are omitted here due to the space limitation.}

\begin{figure}[htbp]
  \centering
  \begin{minipage}[t]{0.45\linewidth}
  \centering
  \includegraphics[scale=1]{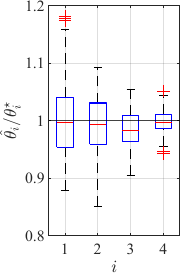}
  \subcaption{Stabilized OE $(\hat{K}=K)$}
  \end{minipage}
  \quad
  \begin{minipage}[t]{0.45\linewidth}
  \centering
  \includegraphics[scale=1]{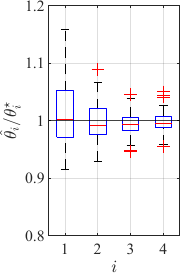}
  \subcaption{Fixed-pole observer model}
  \end{minipage}
  \caption{Distribution of estimates by virtual controller selection (magnetic levitation system)}
  \label{fig:ex:maglev:thhat}
\end{figure}

\rv{Instead, the statistics of the parameter estimates over 100 trials are as shown in Fig.}~\ref{fig:ex:maglev:thhat}\rv{, where the red bar shows the median and the blue rectangles shows the area between the upper quarter and the lower quarter.}
 The bias in the estimates is acceptable compared to the case where $K(p)$ is completely known.

Also, in this example, the estimate is not very sensitive to the arrangement of the poles. 
For example, changing the real part of the poles in $\pob$ that is set to $-3$ from $-0.1$ to $-10$ hardly changes the estimates.

This example suggests that the proposed method is effective for the problem of continuous-time system identification in a closed-loop with unknown controllers, for which no promising method is known.
\rv{The computation time is about 10 seconds for each trial on one AMD Zen 2 core.}

\subsection{Non-stationary data and fixed-pole extended observer model}

Real-world applications often require system identification from time-series data with non-stationary trends.
Prediction error methods based on discrete-time models can deal with this type of data by using models with integrators in the noise part (e.g., ARIX and ARIMAX models).
For continuous-time system identification, the fixed-pole extended observer model in Section~\ref{sec:proposed:exo} can accommodate this type of data.
And, the validity of the method is confirmed here.

\begin{figure}[htbp]
  \centering
  \includegraphics[scale=1.0]{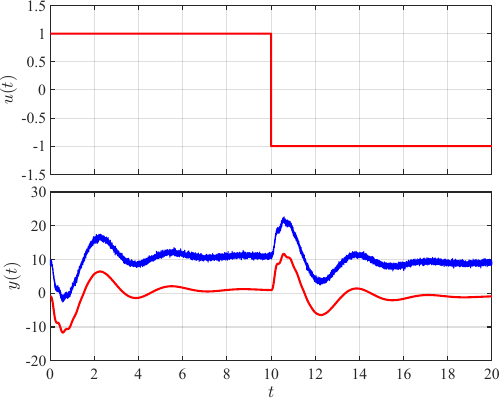}
  \caption{Example of input/output data for identification (Rao–Garnier test system with constant disturbance)}
  \label{fig:ex:rg-c:io}
\end{figure}

In order to make a comparison with the usual OE based continuous-time system identification method, which is not applicable to closed-loop identification of unstable systems, we consider here a stable plant (Rao–Garnier test system \cite{RG2002,Gan15}) described by the following transfer function
\begin{align}
    P(p,\theta) &= \frac{\theta_5 p + \theta_6}{p^4 + \theta_1 p^3 + \theta_2 p^2 + \theta_3 p + \theta_4},\\
    \theta^\star &= [5, 408, 416, 1600, -6400, 1600]^\top,
\end{align}
and data generating system is the one shown in Fig.~\ref{fig:datagen} with $K=0$.
To reproduce the problem in data with non-stationary trends, we set the disturbance to $w(t) = 10$, and the measurement noise $\eta$ is zero-order hold signal of random numbers sampled every $\SI{1}{ms}$ from $\mathcal{N}(0, 0.4^2)$.
An example of I/O data (blue lines) and its noise-free version (red lines) are shown in Fig.~\ref{fig:ex:rg-c:io}.
Note that the red and blue lines for the input $u(t)$ coincide because the plant is in an open-loop, and the disturbance $w(t)$ is not visible in $u(t)$. 
The effect of the constant disturbance appears as an offset in the output $y(t)$.
The sampling interval is $\SI{1}{ms} (=h)$, and the number of data is $\SI{20000}{} (=N)$.

\begin{figure}[htbp]
  \centering
  \begin{minipage}[b]{0.99\linewidth}
  \centering  
  \includegraphics[scale=1]{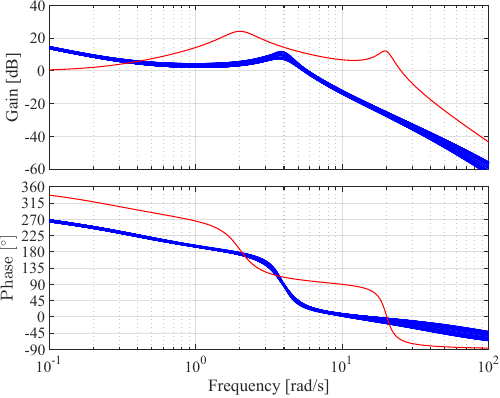}
  \subcaption{Conventional (\texttt{tfest})}
  \label{fig:ex:rg-c:bode:tfest}
  \end{minipage}
  \\[2mm]
  \begin{minipage}[b]{0.99\linewidth}
  \centering  
  \includegraphics[scale=1]{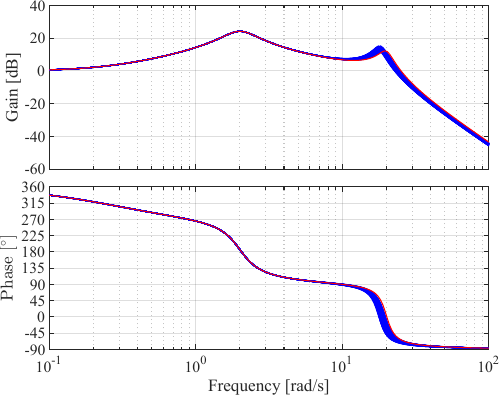}
  \subcaption{Stabilized OE (fixed-pole extended observer model)}
  \label{fig:ex:rg-c:bode:soe_exo}
  \end{minipage}
  \caption{Frequency response of model (Rao–Garnier test system with constant disturbance)}
  \label{fig:ex:rg-c:bode}
\end{figure}

First, continuous-time models obtained with the \texttt{tfest} function in MATLAB System Identification Toolbox is shown in Fig.~\ref{fig:ex:rg-c:bode:tfest} for comparison.
It can be seen that inappropriate models are obtained, although the conditions are sufficient for identification if there are no constant disturbances.

On the other hand, the results obtained by the stabilized OE, where the virtual controller forms the fixed-pole extended observer with $\pob=\left\{-3, -3 \pm i, -3 \pm \frac{i}{2} \right\}$, are shown in Fig.~\ref{fig:ex:rg-c:bode:soe_exo}.
The figure shows that although some bias can be observed, appropriate models have been obtained.

\begin{figure}[htbp]
  \centering
  \begin{minipage}[t]{0.47\linewidth}
  \centering
  \includegraphics[scale=1]{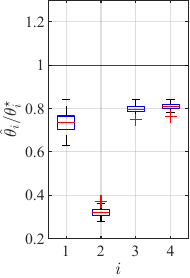}
  \subcaption{Fixed-pole observer model}
    \label{fig:ex:maglev-c:thhat:ob}
  \end{minipage}
  \quad
  \begin{minipage}[t]{0.47\linewidth}
  \centering
  \includegraphics[scale=1]{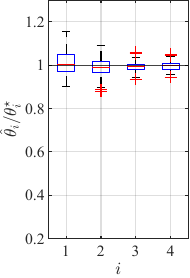}
  \subcaption{Fixed-pole extended observer model}
\label{fig:ex:maglev-c:thhat:exob}
  \end{minipage}
  \caption{Distribution of estimates by virtual controller selection (magnetic levitation system with constant disturbance)}
  \label{fig:ex:maglev-c:thhat}
\end{figure}

In addition, to confirm that this approach is also valid for closed-loop identification of unstable plants, we show the results for the example in Section~\ref{sec:ex:maglev} when the disturbance is set to $w(t) = 1$.
Fig.~\ref{fig:ex:maglev-c:thhat:ob} shows the results with the fixed-pole observer model in Section~\ref{sec:ex:maglev}, and Fig.~\ref{fig:ex:maglev-c:thhat:exob} shows the results from the fixed-pole extended observer model with $\pob=\left\{-3\pm i, -3 \pm \frac{i}{2}\right\}$.
As can be seen in the figure, there is a strong bias in the estimates from the model with no input correction term, while there is no apparent bias in the estimates from the extended observer model, confirming that the proposed method works as intended in this case as well.

\section{CONCLUSIONS}

This paper has discussed closed-loop identification to obtain a continuous-time model from I/O data alone and proposed a new method. 
The method uses a new continuous-time prediction model named pole-fixed observer, which is a reasonable continuous-time version of the innovation model in discrete-time systems.
The proposed method is also a variant of the stabilized output error method which stabilizes the prediction model by a virtual controller, and has good usability features in common with the popular output error method, such as robustness to noise with complex dynamics and applicability to a wide class of models.
An important feature of the proposed method is its applicability to unstable plants in closed-loop without any knowledge of controllers, for which no promising method is yet known.
\rv{In addition, it is shown that the proposed method is a convex optimization problem.}
Since the continuous-time model has advantages such as freedom in selecting sampling intervals and easy access to physical knowledge, it is significant that this type of system identification can be performed based on a continuous-time model.
While there may be many useful variants of virtual controllers that stabilize the predictive model, we present here a case for constructing a fixed-pole expanded observer that can deal with data with non-stationary trends.

Future work includes pursuing further variations of the virtual controller and validating the idea of the proposed method in nonlinear system identification.

\bibliographystyle{IEEEtran}
\bibliography{IEEEabrv,ref}

\end{document}